\documentclass[runningheads]{llncs}
\usepackage{graphicx}
\begin{document}
\title{Distributed Learning for Melanoma Classification using Personal Health Train}
\subtitle{TMF Workshop Usecase Implementation, Sep 2020}
%
%
\author{Yongli Mou\inst{1} \and Sascha Welten\inst{1} \and Yeliz Ucer Yediel\inst{2} \and Toralf Kirsten\inst{3} \and Oya Deniz Beyan\inst{1,2}}

%
\institute{RWTH Aachen University, Ahornstr. 55, 52056 Aachen, Germany 
\and
Fraunhofer FIT, Schloss Birlinghoven, 53757 Sankt Augustin, Germany
\and
Hochschule Mittweida, Germany
}

\maketitle              


\section{Introduction}

Skin cancer is the most common cancer type. Usually, patients with suspicion of cancer are treated by doctors without any aided visual inspection. At this point, dermoscopy has become a suitable tool to support physicians in their decision making. However, clinicians need years of expertise to classify possibly malicious skin lesions correctly\footnote{https://www.isic-archive.com/}.

Therefore, research has applied image processing and analysis tools to improve the treatment process. In order to perform image analysis and train a model on dermoscopic images data needs to be centralised. Nevertheless, data centralisation does not often comply with local data protection regulations due to its sensitive nature and due to the loss of sovereignty if data providers allow unlimited access to the data.

A method to circumvent all privacy-related challenges of data centralisation are Distributed Analytics (DA) approaches, which bring the analysis to the data instead of vice versa \cite{Beyan:2020}. This paradigm shift enables data analyses - in our case, image analysis - with data remaining inside institutional borders, i.e., the origin. In this documentation, we describe a straightforward use case including a model training for skin lesion classification based on decentralised data. 

The remainder of this document is structured as follows. The next chapter presents a brief summary of state-of-the-art DA approaches. Section 3 describes our (simulated) scenario of conducting DA with skin lesion data. Finally, section 4 gives insights about the quality of our analysis.

\section{Related Work: Multi-Institutional Deep Learning}

\begin{figure}
\includegraphics[width=\textwidth]{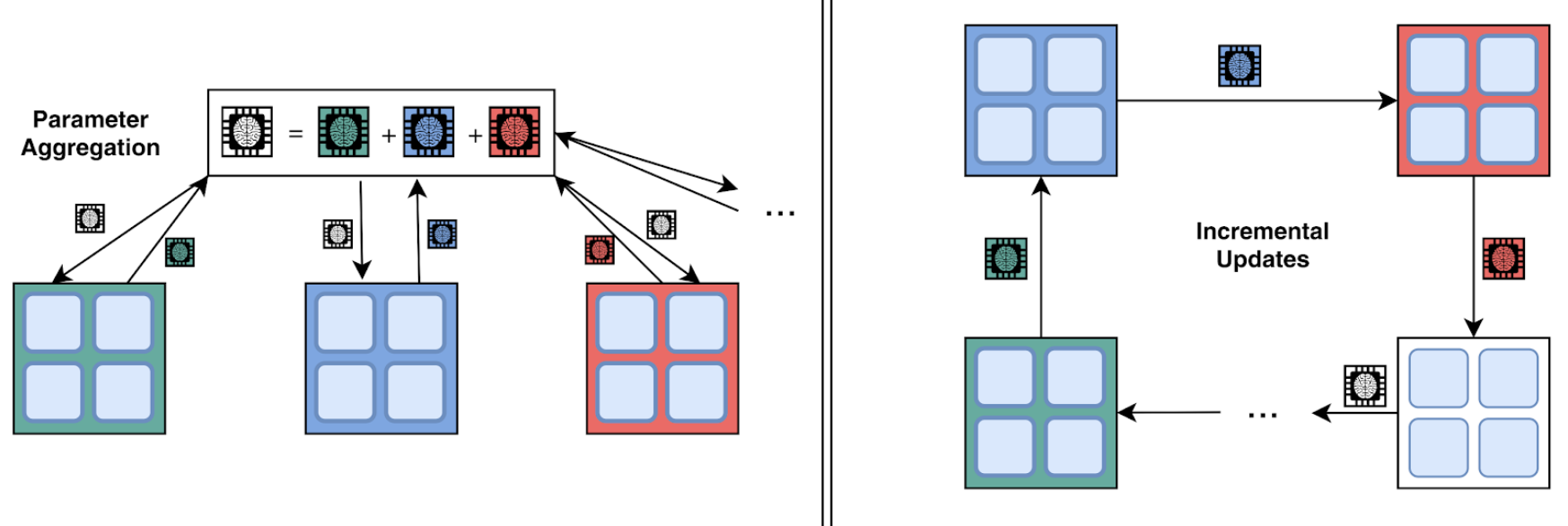}
    \caption{FL (left) and IIL/WT (right). In a DA setting, we usually face two policies for performing an analytical task. The parallel approach, FL, distributes replica of the analysis task to the data providers. The data provider executes the task, e.g., model training, and returns the results to the central component, which aggregates the results. If necessary, the distribution process is repeated - based on the aggregated results. An alternative is the sequential execution of the task following an incremental result update policy. In this scenario, the analytical task is step-wisely circulated inside of a network of data providers. If the analytical task periodically \textit{re-visits} the data providers, it is denoted as cyclic IIL/WT.} 
\label{fig01}
\end{figure}

Several approaches and infrastructures have been developed to perform statistical analyses or model training in terms of decentralised data \cite{Chang2018DistributedDL}\cite{dasbulk}\cite{mcmahan}\cite{sheller}\cite{su_experiments_2015}\cite{sucommunication-efficient}. These approaches follow two basic paradigms of bringing the algorithm to the data. The first paradigm is the parallel execution of the analysis task, which is often referred as to Federated Learning (FL) in the literature \cite{mcmahan}\cite{su_experiments_2015}. Furthermore, there is a successive execution of the analysis tasks, also known as Institutional Incremental Learning (IIL) or Weight Transfer (WT) \cite{Chang2018DistributedDL}\cite{sheller}. An overview of the two paradigms is given in Figure \ref{fig01}.

Since the communication only consists of the results of the analytical task, data never leaves its origin and stays inside of the institutional borders. Due to this circumstance, DA architectures propose a solution for preserving data privacy and comply with the above-mentioned existing regulations.

One platform for DA is the so-called PHT infrastructure \cite{Beyan:2020}\cite{shi_distributed_2019}\cite{Sun:2019}. The PHT originates from an analogy from the real world. The infrastructure reminds of a railway system including trains and stations. The train uses the network to visit different stations to transport, e.g., several goods. Adapting this concept to the PHT ecosystem, we can draw the following similarities. The train encapsulates an analytical task, which is represented by the good in the analogy. The data provider takes over the role of a reachable Station, which can be accessed by the Train.  Further, the Station executes the task,  which processes the available data. In general, the PHT processes the analytical task incrementally by visiting each station one by one. However, parallel executions are also possible by integrating a central aggregation component. Therefore, our work considers both policies the parallel and the incremental approach. Note that we denote Station and data provider, as well as, Train and analytical task interchangeably. Other PHT architectures,  which have been developed by the scientific community, have been applied to several use cases \cite{Beyan:2020}\cite{Sun:2019},\cite{shi_distributed_2019},\cite{deist_distributed_2020},\cite{jochems_distributed_2016},\cite{jochems_developing_2017}.

Deist et al. have conducted DA experiments based on decentralised lung cancer patient data \cite{deist_distributed_2020}. They have been able to perform several analyses and reveal insights from the data without data centralisation. 
Further, Shi et al. have applied the PHT approach to facilitate the provision of radiomics data for DA \cite{shi_distributed_2019}. They emphasise the tremendous need of such a distributed infrastructure.

In this documentation, we follow a similar DA approach using the PHT implementation of the RWTH Aachen, FIT Fraunhofer, and HS Mittweida. The concrete experiment setup is presented in the following section.

\section{Experiments}

In this section, we introduce the dataset on which we conduct the experiments and the experimental setups.

\subsection{Dataset}

We conduct the experiments on the Skin Lesion Images for Melanoma Classification from ISIC 2019 Challenge (International Skin Imaging Collaboration). The ISIC 2019 dataset  is aggregated from three data provenance. i.e., BCN\_20000 dataset \cite{combalia2019bcn20000}, MSK dataset \cite{codella2018skin} and HAM10000 dataset \cite{tschandl2018ham10000} and is used for the purposes of clinical training and supporting technical research toward automated algorithmic analysis. 

\begin{figure}
\includegraphics[width=\textwidth]{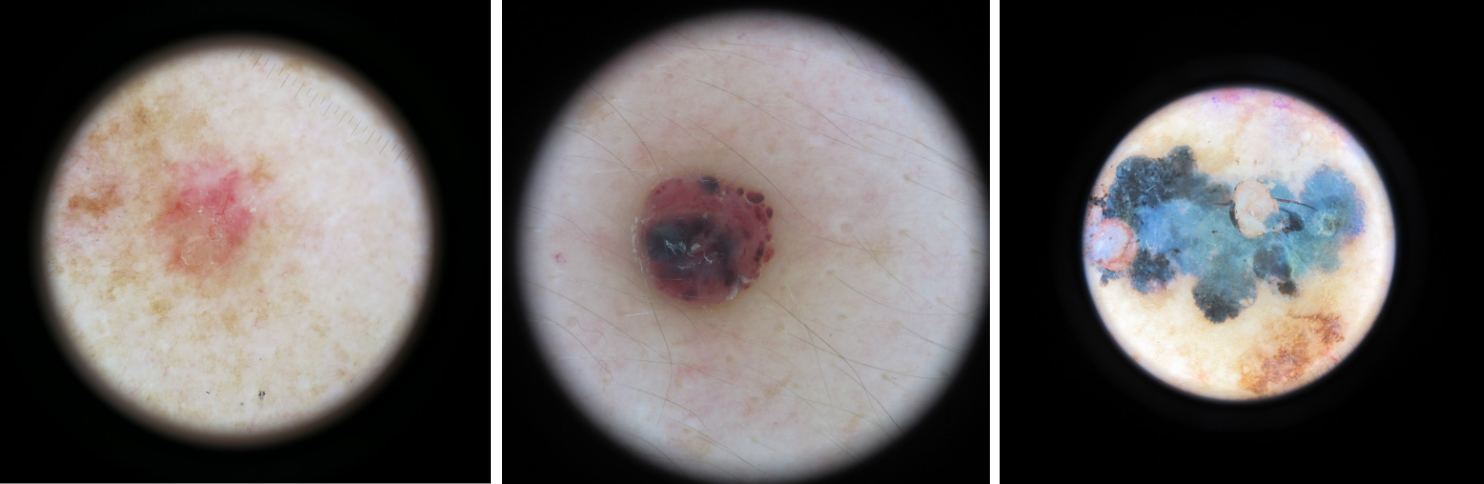}
\caption{Dermoscopic images: (left) basal cell carcinoma (BCC); (center) vascular lesion (VASC) and (right) melanoma (MEL)} \label{fig02}
\end{figure}

The obtained dermoscopic images are in different resolutions. The images from HAM10000 dataset are in size of 600$\times$450, those from BCN\_20000 dataset are in size of 1024$\times$1024 and MSK dataset contains images with various sizes. The official training dataset contains 25331 dermoscopic images and corresponding labels and metadata. The images are classified among nine different diagnostic categories, i.e., melanoma (MEL), melanocytic nevus (NV), basal cell carcinoma (BCC), actinic keratosis (AK), benign keratosis (BKL), dermatofibroma (DF), vascular lesion (VASC) and squamous cell carcinoma (SCC). Fig.~\ref{fig02} shows some example dermoscopic images.

Moreover, each image is described with some patient meta information, e.g., the approximate age, the anatomical site and the sex. However, the meta information is only partially complete. 

Figure 3 displays the class distribution of the ISIC 2019 dataset in percentage. From the histogram we can note that the diagnostic category melanocytic nevus (NV) is the majority among all classes covering about half of the dataset, the other classes melanoma (MEL), basal cell carcinoma (BCC) and benign keratosis (BKL) occupy 18\%, 13\% and 10\% respectively and the rest classes actinic keratosis (AK), dermatofibroma (DF), vascular lesion (VASC) and squamous cell carcinoma (SCC) take up together around 8\%. For developing robust models, the influence of class imbalance should be taken into account while training and evaluation.

\begin{figure}
\includegraphics[width=.9\textwidth]{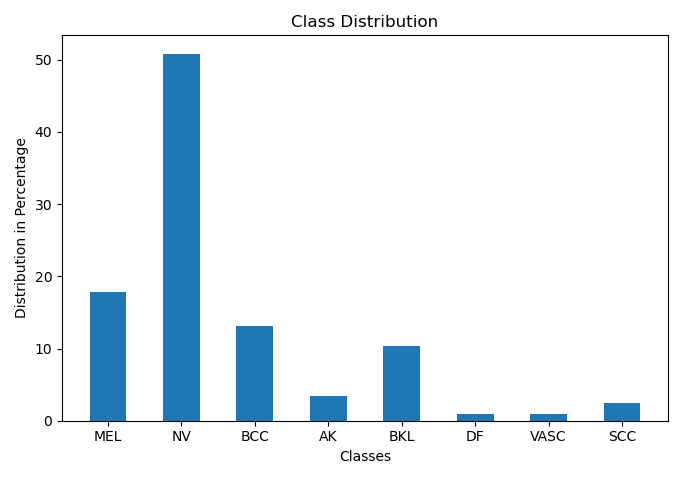}
\caption{Class distribution of the multi-class dataset. Overall, we have eight different cancer types, which we want to classify. Most common is the class NV ($50$\%). DF is least present (10\%).} 
\label{fig03}
\end{figure}

\subsection{Data Management and Access}

In this section, we present the data management and data access at each station in more detail. Figure 3 gives an overview of the data storage and retrieval. 

As described above, we use two data types for our experiment. For each data type, we use a seperate storage system. For the mentioned patient metadata, we use a Fast Healthcare Interoperability Resources (FHIR) server. FHIR is a standard published by Health Level Seven (HL7) International health-care standards organization aiming for health care data exchange. The standard describes data formats and elements and APIs for exchanging electronic health records. In our scenario, we use a Blaze-Server to provide the FHIR interfaces. We use three FHIR resources. We model the ISIC Challenge as a so-called ImageStudy resource to give information under which circumstances the image has been created. Each patient described in the metadata is saved in a Patient resource capturing information about age or height for example. Each Patient resource has a link to a Media resource saving a link to the actual dermoscopy image. More precisely, each Media resource captures a reference to the location of the patient’s skin lesion image, which is stored at a different location.

Therefore, the clinical images (second data type), are stored in a standalone object storage and management system. In our scenario, the MinIO file storage system provides the desired functionality of storing images at predefined locations.

Based on this preliminary work, the Train encapsulating the analytic task is accessing the data as follows. First, the Train image is pulled. During the container instantiation, the Station admin provides connection information to the container. After executing the container, the analytic task queries the Patient resources of the Image Study to obtain the image references (URLs) stored in the Media resource. The analytic task downloads the images from the MinIO server based on the given URLs and performs the image analysis, i.e., the model training. Finally, the model updates are saved (Container commit) and the updated Train image is sent back to the Train repository.

\begin{figure}
\includegraphics[width=\textwidth]{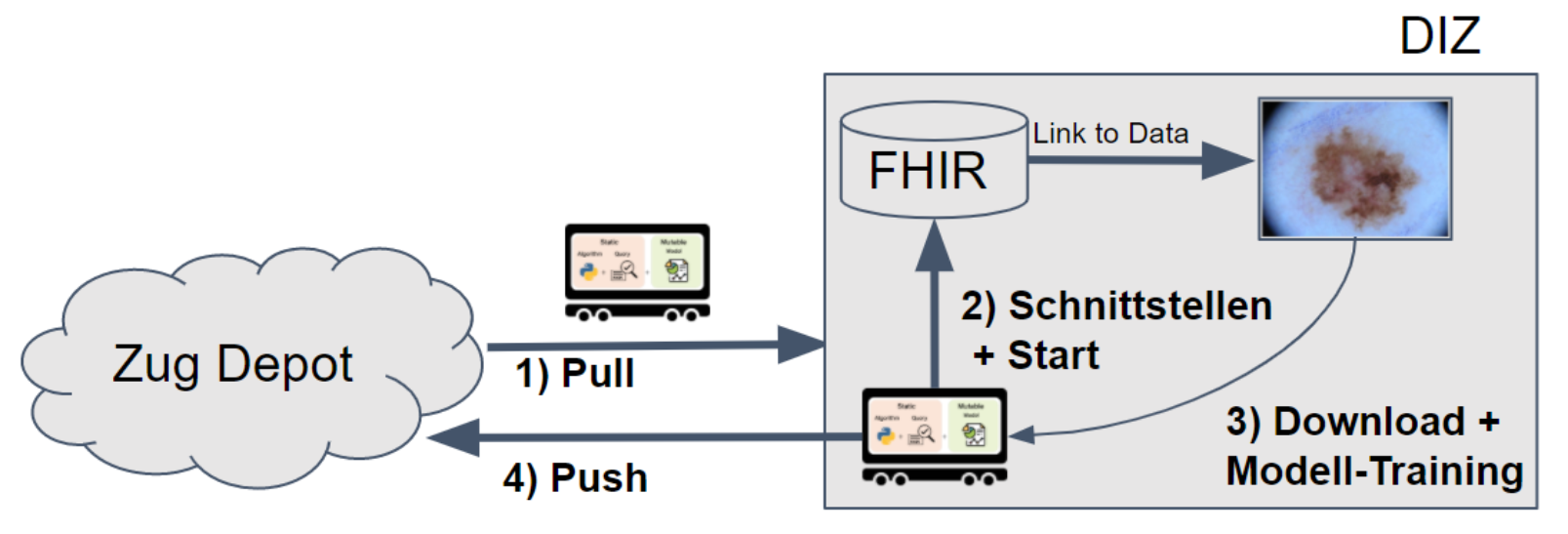}
\caption{Data Management and Access at each Station. Patient's data is stored in a FHIR instance as a single point of contact for the data queries. After the Train is pulled, the Train queries the data. Each patient record provides a reference to the actual image location. In our case, this is a MinIO server, which allows the storage of the dermoscopy images. The Train downloads the images from the corresponding locations and starts the training routing. Finally, the Train is pushed back.} \label{fig04}
\end{figure}

\subsection{Experimental Setups}
In this section, we present the experimental setups, including data distribution, image pre-processing and data augmentation as well as models we used in the experiments.

In the experiments, we demonstrate the institutional incremental learning where the data are distributed on three physically separated locations, i.e, two workstations in Aachen and one in Mittweida. 

The official training dataset is split into subsets for the training, the internal evaluation and the final evaluation. The test dataset for the final evaluation contains 5066 images covering 20\% of the ISIC 2019 dataset. The other 80\% images are split into three subsets and distributed among three stations, i.e., each station obtains 6755 images. 80\% of images on each station are used as training dataset and the other 20\% are used as validation dataset for internal evaluation. Figure 4 shows the class distribution of training and validation datasets on each station. It is observed that the classes are independent and identically distributed (i.i.d.) among stations, which is the ideal situation. In practice, we need to develop robust algorithms or models to address Non-i.i.d. data. For comparison, we train the models using the centralized approach. The training datasets on each station are collected in the central location as well as the validation datasets for internal evaluation.

\begin{figure}
\includegraphics[width=\textwidth]{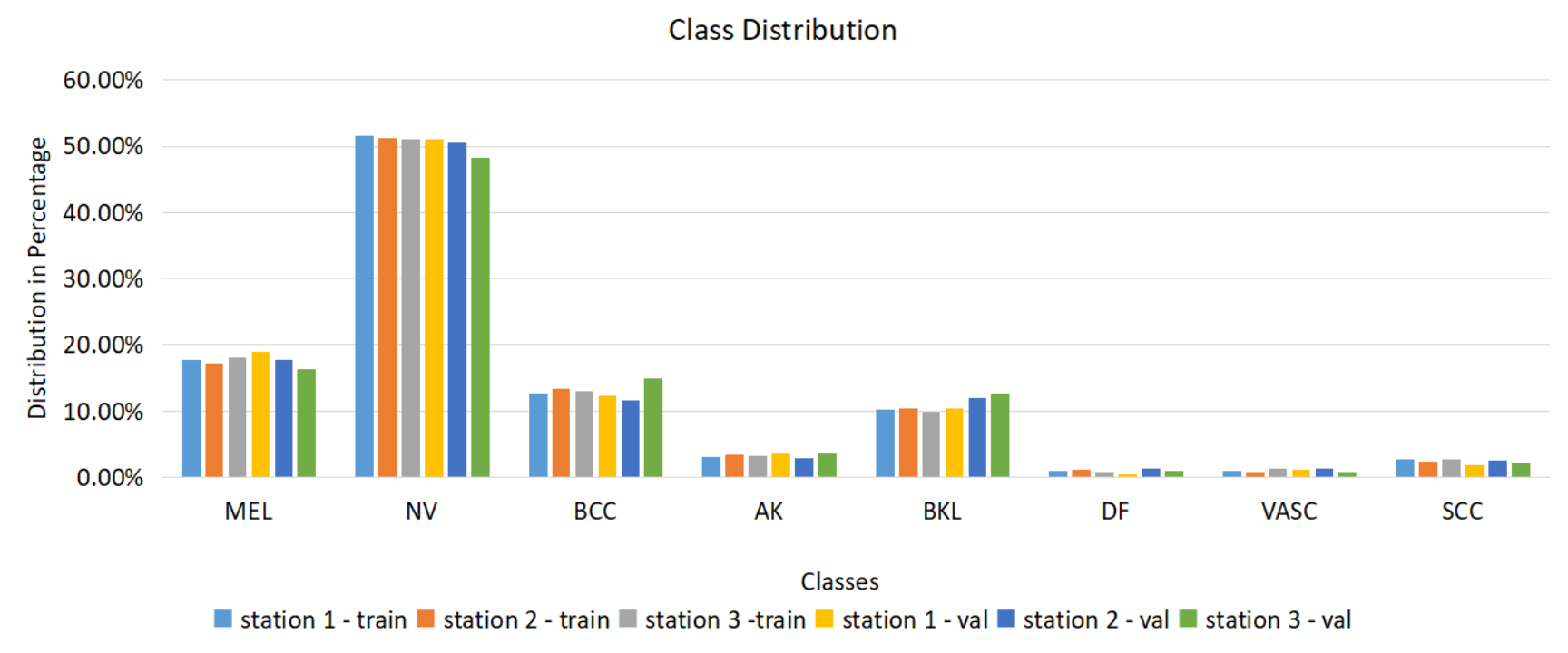}
\caption{Station-wise distribution of the initial dataset. To enable model training, we equally and artificially distribute the data among our three stations. This results in three subsets of the main dataset showing a similar class distribution.} \label{fig05}
\end{figure}

As a first step, we use a center cropping strategy to deal with the problem of various image resolutions. The cropped images are then resized in the resolution of 256×256 as the fixed size fed into convolutional neural networks. Next, we apply the geometric and photometric data augmentation techniques as regularization to reduce overfitting, for example, random horizontal flip, random vertical flip and color jittering of brightness, contrast, saturation and hue.

The experiments are conducted on the representative image classification architecture: ResNet, that have been pretrained on the ImageNet dataset. This ResNet model family contains five different models with different depths, namely, ResNet-18, 34, 50, 101 and 152. For the further comprehensive experiments to proof the robust of distributed approaches, the other models will be used.  

We implement our algorithm with Pytorch deep learning framework. Training is performed on RWTH Cluster using NVIDIA Tesla P100 SXM2 graphics cards. We train the model for 40 epochs using Adam as the optimizer. The initial learning rate is 0.0001, the weight decay is 0.0005 and the batch size is 16. 


\section{Results}
In this section, we present the experimental results of our approach for distributed deep learning on the ISIC 2019 dataset. 

As for the metrics, we use the mean accuracy for the internal evaluation during the training process and the mean accuracy \ref{acc} and the mean recall \ref{recall} (as known as  mean sensitivity) for the final evaluation on the test dataset. The formulations of metrics are listed in the following, where $C$ is the number of classes, $TP_i$ is the number of true positive of class $i$, $FN_i$ is the number of false negative of class $i$, $FP_i$ is the number of false positive of class $i$ and $TN_i$ is the number of true negative of class $i$. In addition, the cross entropy loss is also recorded as an indicator of training.

\begin{equation}
    Recall = \frac{1}{C} \sum^{C}_{i=1} \frac{TP_i}{TP_i+FN_i}
    \label{recall}
\end{equation}

\begin{equation}
    Accuracy = \frac{1}{C} \sum^{C}_{i=1} \frac{TP_i}{TP_i+FN_i+FP_i+TN_i}
    \label{acc}
\end{equation}

\begin{figure}
\includegraphics[width=.8\textwidth]{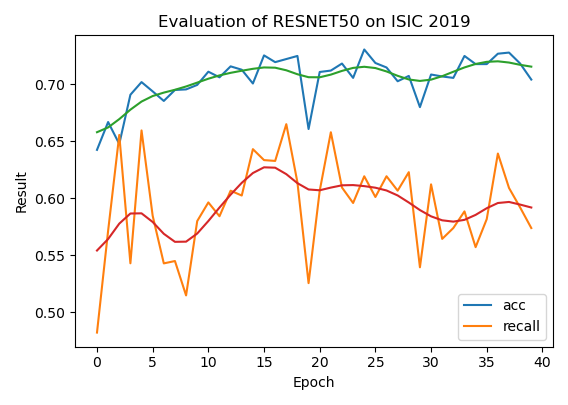}
\caption{Results of the centralized approach.} \label{fig06}
\end{figure}

\begin{figure}
\includegraphics[width=\textwidth]{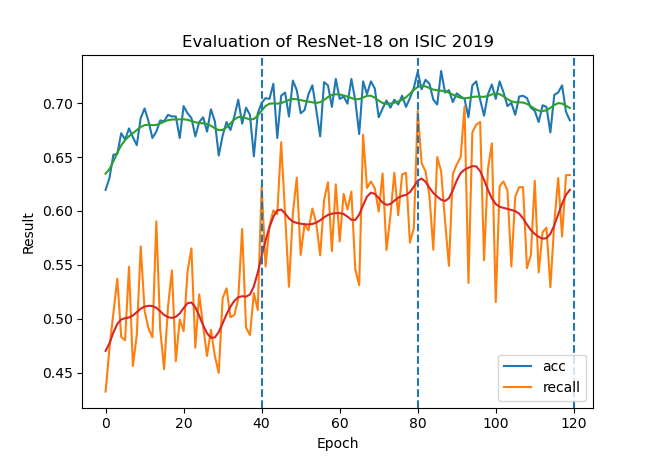}
\caption{Results of the distributed approach.} \label{fig07}
\end{figure}

Fig.~\ref{fig06} and Fig.~\ref{fig07} shows the overall accuracy and recall on internal validation set for centralized approach and distributed approach respectively. The graph shows that there has been a trend of the increasing of overall accuracy in both approaches. The mean recall shows also a rising trend but not stable. The results for final evaluation on test dataset mentioned above with centralized and distributed approaches are shown in Table~\ref{test-result}.

\begin{table}[]
\centering
\begin{tabular}{|l|l|l|}
\hline
Metrics     & Mean Accuracy & Mean Recall \\ \hline
Centralized Approach&       75.40        &        69.22     \\ \hline
Distributed Approach&       71.83        &        63.35      \\ \hline
\end{tabular}
\caption{Final evaluation of ResNet-18 on test dataset}
\label{test-result}
\end{table}

\begin{figure}
\includegraphics[width=\textwidth]{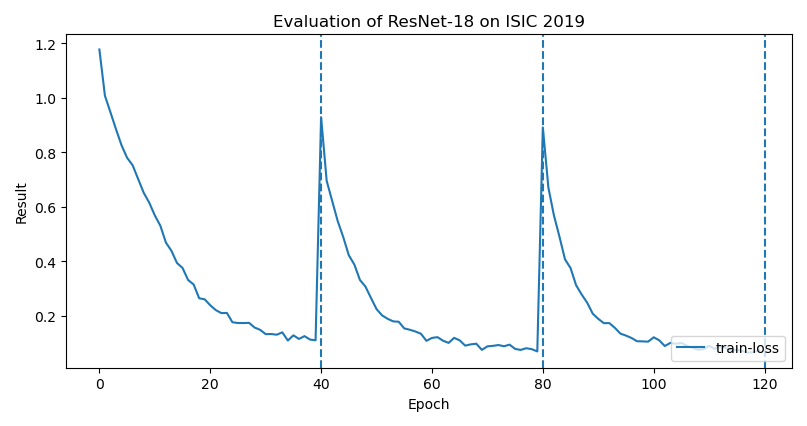}
\caption{Training process of our model Train. We depict the training progress according to the train loss. After every 40th epoch, we send the Train to the next Station. It is notably, that after every Station-hop, the train loss increases. This phenomenon is related to catastrophic forgetting - unlearning of already learned features.} \label{fig08}
\end{figure}

\section{Conclusion}

In this documentation, we presented our methodology to conduct DA using a PHT architecture developed by the institutions RWTH Aachen, FIT Fraunhofer, and HS Mittweida. We presented an exemplary image analysis use case. Precisely, we aim at performing a distributed model training on skin lesion images. The final model should be capable to classify a held-out test set and show a comparable accuracy to the centralised approach. We artificially distribute the data to three Germany-wide geographically-separated Stations and create a Train suitable for such an analysis task. After visiting each data provider, we obtain a model having a similar but slightly lower performance as the centrally trained model. However, we have shown that our architecture, presented in the demo, enables a complex model training on decentralised data.

%
%
\bibliographystyle{splncs04}
\bibliography{ref}

\begin{thebibliography}{10}
\providecommand{\url}[1]{\texttt{#1}}
\providecommand{\urlprefix}{URL }
\providecommand{\doi}[1]{https://doi.org/#1}

\bibitem{Beyan:2020}
Beyan, O., Choudhury, A., van Soest, J., Kohlbacher, O., Zimmermann, L.,
  Stenzhorn, H., Karim, M.R., Dumontier, M., Decker, S., da~Silva~Santos,
  L.O.B., Dekker, A.: Distributed analytics on sensitive medical data: The
  personal health train. Data Intelligence  \textbf{2}(1-2),  96--107 (2020).
  \doi{10.1162/dint\_a\_00032}

\bibitem{Chang2018DistributedDL}
Chang, K., Balachandar, N., Lam, C.K., Yi, D., Brown, J.M., Beers, A., Rosen,
  B.R., Rubin, D.L., Kalpathy-Cramer, J.: Distributed deep learning networks
  among institutions for medical imaging. In: JAMIA (2018)

\bibitem{codella2018skin}
Codella, N.C., Gutman, D., Celebi, M.E., Helba, B., Marchetti, M.A., Dusza,
  S.W., Kalloo, A., Liopyris, K., Mishra, N., Kittler, H., et~al.: Skin lesion
  analysis toward melanoma detection: A challenge at the 2017 international
  symposium on biomedical imaging (isbi), hosted by the international skin
  imaging collaboration (isic). In: 2018 IEEE 15th International Symposium on
  Biomedical Imaging (ISBI 2018). pp. 168--172. IEEE (2018)

\bibitem{combalia2019bcn20000}
Combalia, M., Codella, N.C., Rotemberg, V., Helba, B., Vilaplana, V., Reiter,
  O., Carrera, C., Barreiro, A., Halpern, A.C., Puig, S., et~al.: Bcn20000:
  Dermoscopic lesions in the wild. arXiv preprint arXiv:1908.02288  (2019)

\bibitem{dasbulk}
Das, A., Upadhyaya, I., Meng, X., Talwalkar, A.: Collaborative filtering as a
  case-study for model parallelism on bulk synchronous systems. In: Proceedings
  of the 2017 ACM on Conference on Information and Knowledge Management. p.
  969–977. CIKM ’17, Association for Computing Machinery, New York, NY, USA
  (2017). \doi{10.1145/3132847.3132862},
  \url{https://doi.org/10.1145/3132847.3132862}

\bibitem{deist_distributed_2020}
Deist, T.M., Dankers, F.J.W.M., Ojha, P., Scott~Marshall, M., Janssen, T.,
  Faivre-Finn, C., Masciocchi, C., Valentini, V., Wang, J., Chen, J., Zhang,
  Z., Spezi, E., Button, M., Jan~Nuyttens, J., Vernhout, R., van Soest, J.,
  Jochems, A., Monshouwer, R., Bussink, J., Price, G., Lambin, P., Dekker, A.:
  Distributed learning on 20 000+ lung cancer patients – {The} {Personal}
  {Health} {Train}. Radiotherapy and Oncology  \textbf{144},  189--200 (Mar
  2020). \doi{10.1016/j.radonc.2019.11.019},
  \url{http://www.sciencedirect.com/science/article/pii/S0167814019334899}

\bibitem{jochems_developing_2017}
Jochems, A., Deist, T.M., El~Naqa, I., Kessler, M., Mayo, C., Reeves, J.,
  Jolly, S., Matuszak, M., Ten~Haken, R., van Soest, J., Oberije, C.,
  Faivre-Finn, C., Price, G., de~Ruysscher, D., Lambin, P., Dekker, A.:
  Developing and {Validating} a {Survival} {Prediction} {Model} for {NSCLC}
  {Patients} {Through} {Distributed} {Learning} {Across} 3 {Countries}.
  International Journal of Radiation Oncology, Biology, Physics
  \textbf{99}(2),  344--352 (Oct 2017). \doi{10.1016/j.ijrobp.2017.04.021},
  \url{https://doi.org/10.1016/j.ijrobp.2017.04.021}, publisher: Elsevier

\bibitem{jochems_distributed_2016}
Jochems, A., Deist, T.M., van Soest, J., Eble, M., Bulens, P., Coucke, P.,
  Dries, W., Lambin, P., Dekker, A.: Distributed learning: {Developing} a
  predictive model based on data from multiple hospitals without data leaving
  the hospital – {A} real life proof of concept. Radiotherapy and Oncology
  \textbf{121}(3),  459--467 (Dec 2016). \doi{10.1016/j.radonc.2016.10.002},
  \url{https://doi.org/10.1016/j.radonc.2016.10.002}, publisher: Elsevier

\bibitem{mcmahan}
McMahan, H.B., Moore, E., Ramage, D., Hampson, S., y~Arcas, B.A.:
  Communication-efficient learning of deep networks from decentralized data
  (2016)

\bibitem{sheller}
Sheller, M.J., Reina, G.A., Edwards, B., Martin, J., Bakas, S.:
  Multi-institutional deep learning modeling without sharing patient data: A
  feasibility study on brain tumor segmentation. In: International MICCAI
  Brainlesion Workshop. pp. 92--104. Springer (2018)

\bibitem{shi_distributed_2019}
Shi, Z., Zhovannik, I., Traverso, A., Dankers, F.J.W.M., Deist, T.M.,
  Kalendralis, P., Monshouwer, R., Bussink, J., Fijten, R., Aerts, H.J.W.L.,
  Dekker, A., Wee, L.: Distributed radiomics as a signature validation study
  using the {Personal} {Health} {Train} infrastructure. Scientific Data
  \textbf{6}(1), ~218 (Oct 2019). \doi{10.1038/s41597-019-0241-0},
  \url{https://www.nature.com/articles/s41597-019-0241-0}, number: 1 Publisher:
  Nature Publishing Group

\bibitem{su_experiments_2015}
Su, H., Chen, H.: Experiments on parallel training of deep neural network using
  model averaging. CoRR  \textbf{abs/1507.01239} (2015),
  \url{http://arxiv.org/abs/1507.01239}

\bibitem{sucommunication-efficient}
Su, Y., Lyu, M., King, I.: Communication-efficient distributed deep metric
  learning with hybrid synchronization. In: Proceedings of the 27th ACM
  International Conference on Information and Knowledge Management. p.
  1463–1472. CIKM ’18, Association for Computing Machinery, New York, NY,
  USA (2018). \doi{10.1145/3269206.3271807},
  \url{https://doi.org/10.1145/3269206.3271807}

\bibitem{Sun:2019}
Sun, C., Ippel, L., Van~Soest, J., Wouters, B., Malic, A., Adekunle, O.,
  van~den Berg, B., Mussmann, O., Koster, A., van~der Kallen, C., et~al.: A
  privacy-preserving infrastructure for analyzing personal health data in a
  vertically partitioned scenario. In: MedInfo. pp. 373--377 (2019)

\bibitem{tschandl2018ham10000}
Tschandl, P., Rosendahl, C., Kittler, H.: The ham10000 dataset, a large
  collection of multi-source dermatoscopic images of common pigmented skin
  lesions. Scientific data  \textbf{5},  180161 (2018)

\end{thebibliography}

\end{document}